\documentclass[prb,aps,nobibnotes,twocolumn,superscriptaddress,10pt,longbibliography]{revtex4-1}
\usepackage[utf8]{inputenc}
\usepackage[T1]{fontenc}
\usepackage[
  top=8mm,
  bottom=7mm,
  left=1.5cm,
  right=1.5cm,
  headheight=14pt,
  includehead,includefoot,
  heightrounded, 
]{geometry} 
\usepackage{color}
\usepackage{amsmath}
\usepackage{amsfonts}
\usepackage{hyperref}
\usepackage{comment}
\usepackage{multirow}
\usepackage{amsthm}
\usepackage{bm}
\usepackage[final]{changes}

\usepackage{float}
\usepackage{fancyhdr}
\usepackage{filecontents}
\hypersetup{
  colorlinks = true,
  allcolors= blue,
}
\usepackage[percent]{overpic}
\usepackage{tikz,pgfplots}
\usepackage{blkarray}
\usepackage[normalem]{ulem}
\usepackage{soul}

\usepackage{braket}
\usepackage{amsmath}
\usepackage{todonotes}
\usepackage{siunitx}
\usepackage{chemformula}
\usepackage{tikz}
\usetikzlibrary{shapes,arrows, calc}
\usepackage[T1]{fontenc}

\newcommand{\beq}{\begin{equation}}
\newcommand{\eeq}{\end{equation}}
\newcommand{\eqname}{Eq.}

\DeclareMathOperator{\Tr}{Tr}

\DeclareSIUnit\rydberg{Ry}

\usepackage{etoolbox}

\renewcommand{\figurename}{{\sf\textbf{Figure}}}
\makeatother
\DeclareUnicodeCharacter{300}{à}

\usepackage{setspace}

\definecolor{dgreen}{rgb}{0,0.4,0}
\colorlet{Changes@Color}{red}
\setaddedmarkup{\textcolor{dgreen}{#1}}
\setdeletedmarkup{}

\definechangesauthor[name={Monacelli}, color=red]{Mon}
\setremarkmarkup{(#2)}

\DeclareMathAlphabet{\mathsfit}{\encodingdefault}{\sfdefault}{m}{sl}
\SetMathAlphabet{\mathsfit}{bold}{\encodingdefault}{\sfdefault}{bx}{sl}

\makeatletter
\fancypagestyle{footmark}{
  \fancyfoot[L]{\footmark}
}

\makeatother

\makeatletter
\renewcommand*{\@fnsymbol}[1]{\ifcase#1\else\@arabic{\numexpr#1\relax}\fi}
\makeatother
\makeatletter
\newcommand*{\newbibstartnumber}[1]{%
  \apptocmd{\thebibliography}{%
    \global\c@NAT@ctr #1\relax
    \addtocounter{NAT@ctr}{-1}%
  }{}{}%
}
\makeatother

\makeatletter
\renewcommand\frontmatter@abstractwidth{\dimexpr\textwidth\relax}
\makeatother

\begin{document}
\title{\sf \textbf{\Huge
    Black metal hydrogen above 360 GPa driven by proton quantum fluctuations
}}
\author{\sf {{Lorenzo Monacelli}}}

\thanks{\sf Department of Physics, University ``Sapienza'', Rome, Italy}
\author{\sf {{Ion Errea}}}

\thanks{\sf Fisika Aplikatua 1 Saila, Gipuzkoako Ingeniaritza Eskola, University of the Basque Country (UPV/EHU), Europa Plaza 1, 20018, Donostia/San Sebasti\'an, Spain}
\thanks{\sf \\Centro de F\'isica de Materiales (CSIC-UPV/EHU), Manuel de Lardizabal Pasealekua 5, 20018 Donostia/San Sebasti\'an, Spain}
\thanks{\sf \\Donostia International Physics Center (DIPC), Manuel de Lardizabal Pasealekua 4, 20018 Donostia/San Sebasti\'an, Spain}
\author{\sf {{Matteo Calandra}}}
\thanks{\sf Sorbonne Universit\'e, CNRS, Institut des Nanosciences de Paris, UMR7588, F-75252, Paris, France}
\author{\sf {{Francesco Mauri}}}
\thanks{\sf Department of Physics, University ``Sapienza'', Rome, Italy}



    \maketitle

{\sf\textbf{
    \hspace*{-4.5mm}
Production of metallic hydrogen is one of the top three open quests of physics\cite{GinzburgNobel}. Recent low-temperature experiments\cite{Dias2017,Eremets2019_hydro,Loubeyre2019observation} report different
metallization pressures, varying from 360GPa to 490GPa. In this work, we simulate structural properties,
vibrational Raman, IR and optical spectra of hydrogen phase III accounting for proton quantum effects.
We demonstrate that nuclear quantum fluctuations downshift the vibron frequencies
by 25\%, introduce a broad line-shape in the Raman spectra, and reduce the optical gap by 3eV.
We show that hydrogen metallization occurs at 380GPa in phase III due to band
overlap, in good agreement with transport data\cite{Eremets2019_hydro}.
By simulating the optical properties, we predict this state to be a peculiar black metal, transparent in the IR.
The transparent window closes at 450GPa, but the reflectivity remains low, discarding phase III as the shiny metal observed at 490GPa\cite{Dias2017}.
We predict the conductivity onset to increase by 70GPa and the transparent window to increase by 1.3eV when replacing hydrogen by deuterium at 0K, underlining that metallization is driven by quantum fluctuations and is thus isotope dependent. 
We show how hydrogen acquires metallic features (conductivity and brightness) at different pressures, explaining the apparent contradictions in existing experimental
scenarios\cite{Dias2017,Eremets2019_hydro,Loubeyre2019observation}.
}}

\begin{figure}
    \centering
    \includegraphics[width=\columnwidth]{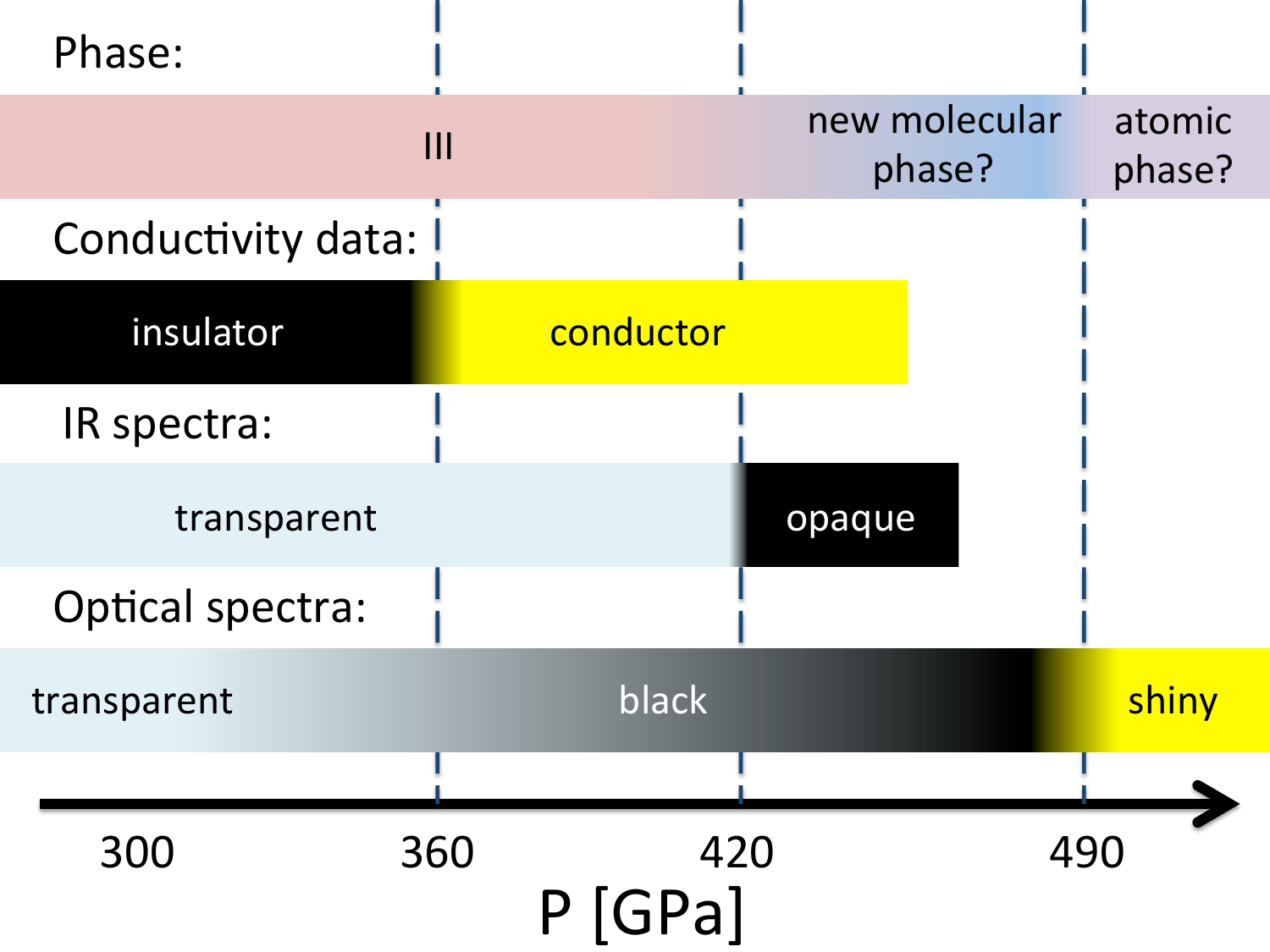}
    \caption{Sketch of the physical properties of low-temperature hydrogen. Optical data and reflectivity measurements\cite{Dias2017} argue that hydrogen is black above \SI{320}{\giga\pascal} and seem to indicate it is shiny at \SI{490}{\giga\pascal}.  Conductivity measurements\cite{Eremets2019_hydro} show that phase III is a conductor above \SI{360}{\giga\pascal}. IR absorption data\cite{Loubeyre2019observation} show a transparency window at low frequency until \SI{420}{\giga\pascal}, where the transmission abruptly drops to zero.
    This is a qualitative sketch, as there are mismatches on the pressure calibration between different experiments\cite{Loubeyre2019observation} and the results are highly debated\cite{Loubeyre2017,Eremets2017,Goncharov2017,SilveraComment2019}.}
    \label{fig:opt:experiments}
\end{figure}

Solid hydrogen at high pressures exhibits a very rich phase diagram with the presence of five different insulating molecular phases, labeled from I to V\cite{mao1994ultrahigh,Howie_2012,DalladaySimpson2016}, before undergoing a transition into the long-sought atomic metallic state proposed by Wigner and Huntington\cite{Wigner1935}, expected to be a room temperature superconductor\cite{AshcroftHydrogen}. The structural characterization of these phases is challenging since both neutron and X-ray scattering require sample sizes compatible with a pressure lower than \SI{250}{\giga\pascal}\cite{Ji2019}. The experimental structural information must be inferred indirectly from vibrational spectroscopy (Raman and IR) and/or optical measurements (transmittance and reflectivity). The experimental difficulties in collecting data and the different probes used have led to apparently contradicting results about the metallization of hydrogen at low temperatures in the solid-state as shown in \figurename~\ref{fig:opt:experiments}. Optical reflectivity measurements suggest metallization of hydrogen in an atomic state at 495 GPa from an unidentified opaque phase\cite{Dias2017}, electrical measurements observe semimetallic behavior in phase III above 360 GPa \cite{Eremets2019_hydro}, while infrared transmission experiments suggest that hydrogen metallizes from phase III after a first-order transition to a metallic phase at 420 GPa\cite{Loubeyre2019observation}.

Numerical \emph{ab initio} simulations play consequently a crucial role in understanding the structures that form the phase diagram of hydrogen. \emph{Ab initio} structural searches successfully elucidated the crystalline morphology of phase II\cite{Kohanoff_1997} and provided good candidates for phases III\cite{Pickard_2007}, IV\cite{Pickard2012}, and V\cite{Monserrat_2018}. However, different theoretical approximations yield to distinct low-energy structures, and, thus, a strong debate on the correct identification of the phases is ongoing\cite{Azadi_2017,azadi2019unconventional}. All structure searches performed so far on hydrogen assume that nuclei are classical particles. The Born-Oppenheimer (BO) energy landscape is explored looking for the global minimum of the energy. However, as hydrogen is the lightest element, its nucleus is subject to huge quantum fluctuations that can largely affect structural properties. For instance, nuclear quantum effects have been shown to completely reshape the BO energy landscape in the hydrogen-rich \ch{H3S}\cite{Errea2015} and \ch{LaH10}\cite{errea2019quantum} compounds, invalidating the results obtained with classical structural searches. Furthermore, many competing structures differ in enthalpy by less than $\SI{1}{\milli\electronvolt}$ per atom in a broad range of pressures\cite{Drummond2015}. This makes the identification of the ground state very sensitive to errors difficult to control, like the choice of a particular exchange-correlation functional in density functional theory (DFT) calculations.


The presence of strong quantum fluctuations is not only crucial for the structures of hydrogen, but it also implies that large anharmonic effects shift and deform the phonon spectral functions observed with Raman and IR probes. Considering that most of the experimental signatures used to distinguish between hydrogen phases rely on the characterization of the \ch{H2} vibron, it is mandatory to perform a fully quantum and anharmonic description of vibrational properties. Classical molecular dynamics simulations of Raman and IR spectra performed so far\cite{Magdau_2013,azadi2019unconventional}, which do consider anharmonicity but not quantum effects, strongly underestimate nuclear fluctuations and may yield to strong errors in vibron energies. In fact, vibrons, with energies above $\SI{3500}{\centi\metre}^{-1}$, require approximately $\SI{5000}{\kelvin}$ to be thermally populated so that their quantum fluctuations are completely missed in any classical molecular dynamics simulation in the solid-state.    


The difficulties in dealing with quantum nuclei lead to the existence of strange twists in the recent theoretical literature: while it is commonly recognized that the Perdew-Burke-Ernzerhof (PBE) exchange-correlation functional of DFT performs poorly for high-pressure hydrogen\cite{Clay_2014,Drummond2015,Azadi_2017,McMinis_2015,Monserrat_2018}, it is still the first choice for comparing vibrational features with experiments\cite{Pickard_2007,Pickard2012,Magdau_2013,Singh_2014,Monserrat_2018}. This is a clear sign that current theoretical methods lack the precision necessary to compare with experiments, and this is compensated with the \emph{ad hoc} choice of the exchange-correlation functional that best fits with experimental data. However, in this way, calculations cannot be predictive and the correct assessment of the crystalline phase is built on the hope of a big error cancellation between exchange-correlation and quantum nuclear effects.


In this work we focus on hydrogen's metallization, exploring the scenario of the direct/indirect band gap closure of phase III, recently observed experimentally\cite{Eremets2019_hydro,Loubeyre2019observation}, as well as its Raman and IR spectra. This phase is stable above $\SI{150}{\giga\pascal}$ at low temperatures (under $\SI{200}{\kelvin}$) and is characterized by the presence of a high infrared activity above $\SI{4000}{\per\centi\metre}$. Both the IR and the Raman vibrons are well-defined peaks that become softer and broader with increasing pressure. To include quantum and anharmonic effects on nuclei we employ the stochastic self-consistent harmonic approximation (SSCHA)\cite{Errea2014,Bianco2017,Monacelli2018A}. The SSCHA performs a quantum variational minimization of the Gibbs free energy assuming that the quantum wave-functions of the nuclei can be represented as a multidimensional Gaussian, parametrized by \emph{centroid} positions, which determine the maximum of the ionic wave-functions, as well as some effective force constants. By evaluating the stress tensor associated with the SSCHA energy\cite{Monacelli2018A}, we relax the crystal structures including also lattice parameters in a fully quantum and anharmonic description. To simulate vibrational properties we exploit the dynamical extension of the SSCHA\cite{Bianco2017}.
Thus, we can compute all the experimentally accessible data in a consistent way both for the structural and spectral properties. 

\begin{figure*}
    \centering
    \includegraphics[width=\textwidth]{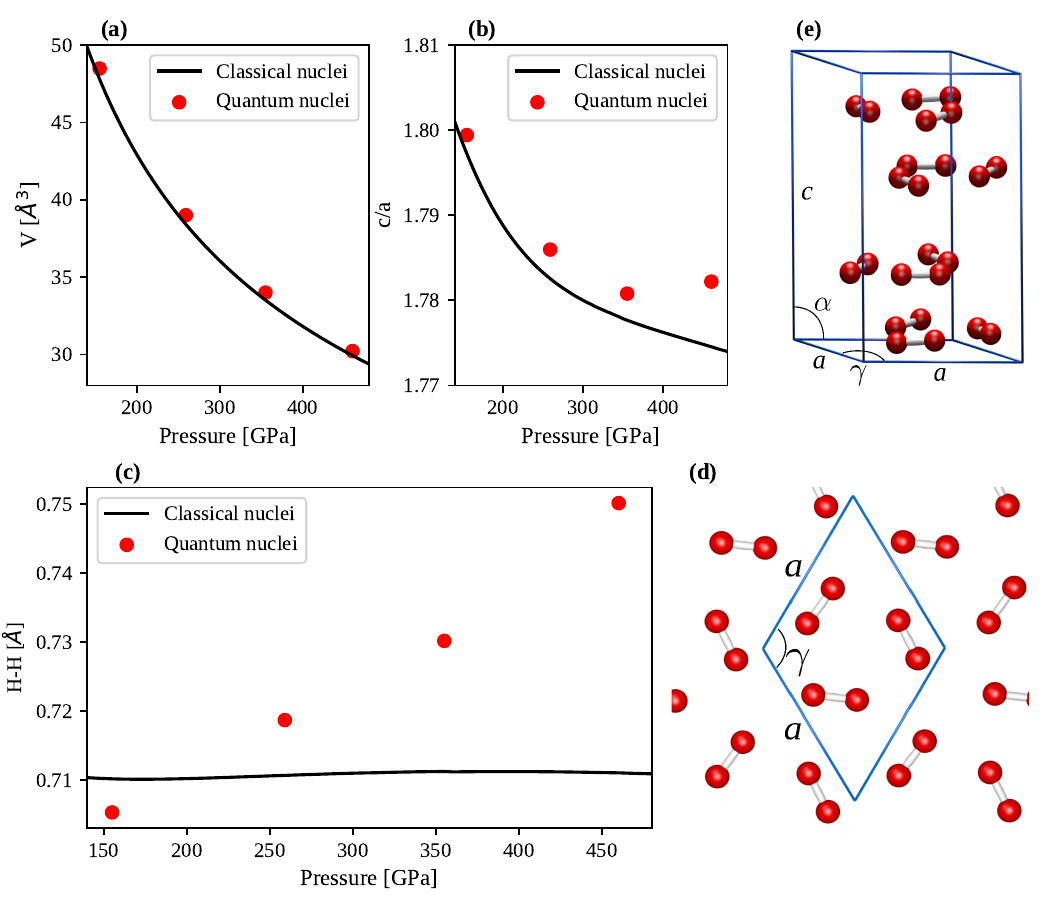}
    \caption{Crystal structure with and without quantum effects of hydrogen phase III (C2/c-24). (\textbf{a}) the equation of state. (\textbf{b}) the $c/a$ ratio as a function of pressure. It quantifies the anisotropy of the quantum contribution to the stress tensor, more pronounced in the $c$ direction (out-of-plane) with respect to the $a$ vector (in-plane). (\textbf{c}) The \ch{H2} bond length as a function of pressure. (\textbf{d}) Top view of a single layer of C2/c-24. (\textbf{e}) Primitive cell of the C2/c-24 phase. }
    \label{fig:hhbond}
\end{figure*}

The most supported candidate for phase III is a base-centered monoclinic structure with C2/c space group and 24 atoms in the primitive unit cell (labeled C2/c-24). The structure consists of four different layers of imperfect hexagonal rings formed by \ch{H2} molecules (see Fig. \ref{fig:hhbond}, panels \textbf{d-e}). Despite its monoclinic character, the cell has a very slight distortion with respect to the hexagonal one ($\alpha = 89.9^o$, $\gamma = 119^o$, the whole structures with quantum effects are reported in the extended data).
We performed a constant pressure quantum relaxation  at \SI{155}{\giga\pascal}, \SI{260}{\giga\pascal}, \SI{355}{\giga\pascal}, and \SI{460}{\giga\pascal} to cover the whole experimental pressure range. We report the structural modification induced by quantum fluctuations in \figurename~\ref{fig:hhbond}.
Quantum fluctuations generate a volume expansion at fixed pressure of about 1.5\% (1\% at \SI{460}{\giga\pascal}), but do not modify qualitatively the equation of state (panel \textbf{a}). The volume expansion is nonisotropic and acts mainly on the out-of-plane lattice parameter $c$, pushing away the layers as pressure is increased (panel \textbf{b}). The most important effect is on the \ch{H2} bond length, which increases up to 6 \% in comparison to the classical result (panel \textbf{c}). A similar extreme stretching was also shown in Cmca-4 hydrogen\cite{Borinaga2016}. The classical treatment of the nuclei completely misses the dependence of the $\ch{H2}$ bond length with pressure in C2/c-24 hydrogen: while the bond length appears to be pressure independent in the classical calculation, it increases with pressure in the quantum calculation, showing the tendency imposed by pressure towards molecular dissociation. Considering that the molecular bond length has a huge impact both on the energy of molecular phases of hydrogen and their vibrational frequencies, no classical calculation is expected to determine accurately the phase diagram and the spectroscopic properties of high-pressure hydrogen. 



\begin{figure*}
    \centering
    \includegraphics[width=\textwidth]{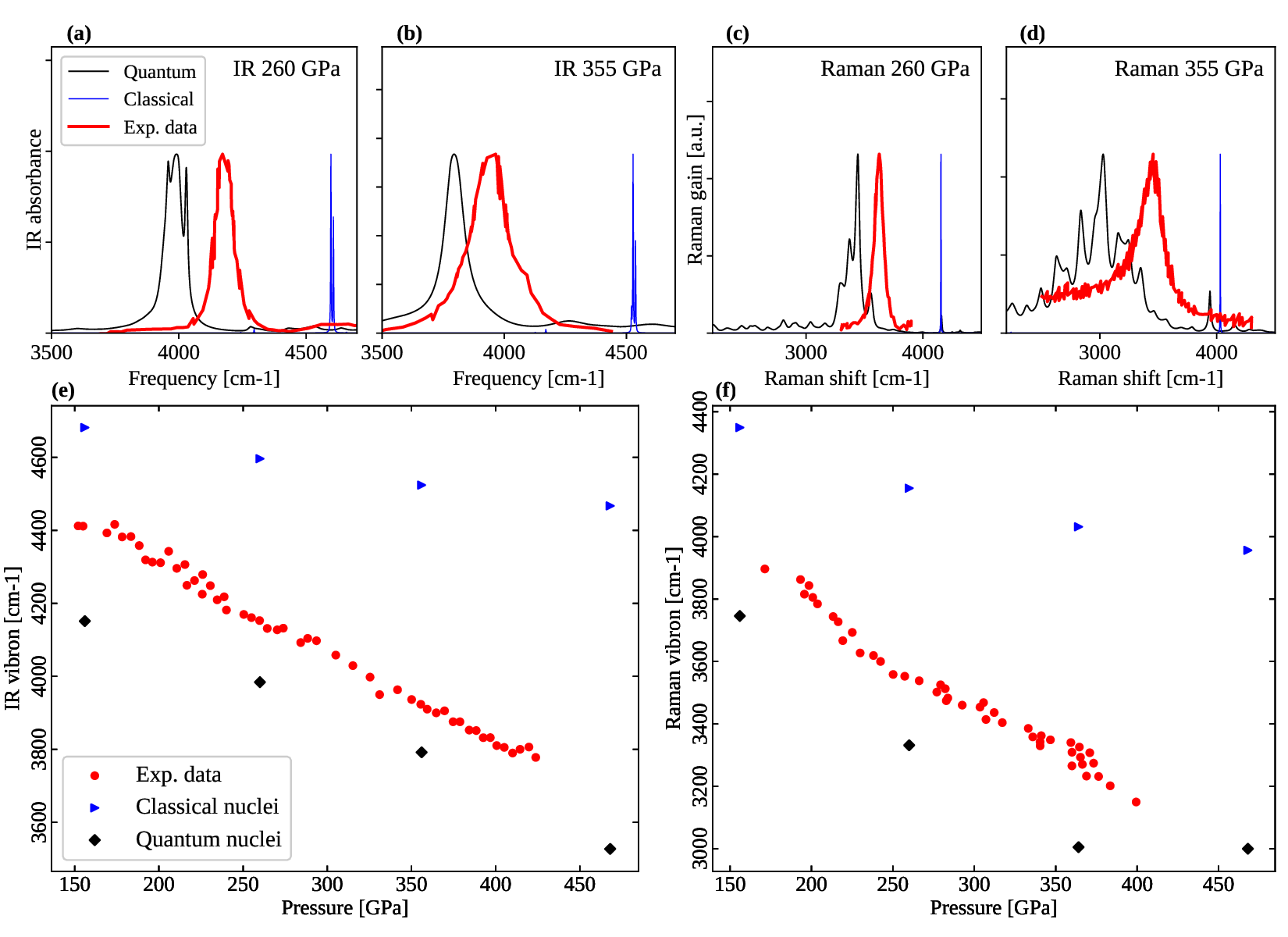}
    \caption{Vibrational spectra at \SI{0}{\kelvin}.  Panels \textbf{(a-b)}: Simulation of the IR absorbance within the harmonic approximation (classical nuclei), full quantum anharmonic theory and experimental data (Ref.\cite{Loubeyre2019observation} at \SI{251}{\giga\pascal} and \SI{351}{\giga\pascal} at  \SI{80}{\kelvin}). A small smearing has been added to the harmonic spectrum for presentation purposes.
    Panel \textbf{(c-d)}:  Simulation of the Raman spectra, experimental data from Ref.\cite{Goncharov2017} (\SI{248}{\giga\pascal} at \SI{140}{\kelvin}) and Ref.\cite{Eremets2019_hydro} (\SI{360}{\giga\pascal} and \SI{100}{\kelvin}). Panel \textbf{(e)}: Position of the IR vibron peak vs pressure. Exp. data from\cite{Loubeyre2019observation} (\SI{80}{\kelvin}). Panel \textbf{(e)}: Raman vibron peak vs pressure. Exp. data from\cite{Eremets2019_hydro}.}
    \label{fig:irramspec}
\end{figure*}

We report in \figurename~\ref{fig:irramspec} the simulation of the Raman and IR vibrons. 
Quantum fluctuations trigger the anharmonicity of the vibron by shifting the position of the peaks and introducing a finite life-time with respect to the harmonic result. The Raman vibron (panels \textbf{c,d}) acquires a very broad linewidth with increasing pressure due to anharmonic phonon-phonon scattering, in good agreement with experiments. On the other side, the IR linewidth does not increase as much as the Raman, and the broadening at \SI{355}{\giga\pascal} is dominated by optical effects. In fact, thanks to the effective charges on the IR vibron, the real part of the dielectric function remains negative above the transverse optical (TO) frequency, causing the sample to absorb in a band between the TO and longitudinal optical (LO) frequencies.
This underlines the importance of electrostatic effects to simulate the vibrational spectrum (see Methods).
The slope of the vibron energy versus pressure increases when quantum effects are considered for both Raman and IR simulations (panels \textbf{e,f}). This is a consequence of the tendency towards dissociation of the \ch{H2} molecules, whose increasing bond-length with pressure is missed by the static theory (\figurename~\ref{fig:hhbond}, panel \textbf{c}). At \SI{300}{\giga\pascal} we predict a slope for the IR vibron of \SI{-1.96}{\per\centi\meter\per\giga\pascal} (harmonic \SI{-0.71}{\per\centi\meter\per\giga\pascal}) increasing the match with the experimental one of \SI{-2.46}{\per\centi\meter\per\giga\pascal}. Also the agreement of the Raman vibron slope is improved: we predict a value of \SI{-3.49}{\per\centi\meter\per\giga\pascal} (harmonic \SI{-1.32}{\per\centi\meter\per\giga\pascal}) versus the experimental one of \SI{-3.02}{\per\centi\meter\per\giga\pascal}.
All the spectra that include quantum fluctuations show a quantitative improvement with respect to the harmonic simulation. The vibron gets much closer to the experimental value, and the line-shape is very well reproduced. Even if we are slightly underestimating the vibrational energy, our results support C2/c-24 as phase III of hydrogen. 
We believe this slight underestimation is a consequence of the choice of the exchange and correlation functional, even if the BLYP functional we use is recognized as the most accurate for predicting energies of high-pressure molecular hydrogen phases\cite{Clay_2014,Drummond2015,Monserrat_2018}. 
The huge peak shift and the change in the slope of the vibron frequency caused by the zero-point motion question the validity of previous calculations that do not include both anharmonicity and quantum nuclear fluctuations\cite{Pickard_2007,Pickard2012,Magdau_2013,Monserrat_2018,Zhang_2018}. 

Even if our quantum anharmonic calculations clearly support C2/c-24 as phase III of hydrogen, it is not obvious if it can explain the apparent contradictions between different experiments on the metallization. To clarify this issue, we compute the optical properties of C2c-24 hydrogen including the electron-phonon interaction beyond perturbation theory. We use the SSCHA quantum wave-function to extract a supercell phonon-distorted configuration to compute the dielectric properties (see Methods). We use supercells with 432 atoms, that are much larger than those used in prior studies (96 atoms)\cite{Rillo_2018,Azadi_2017}. We notice that using a smaller cell introduces spurious metallic states in the gap, overestimating the metallic character. To correct the systematic underestimation of the DFT band gap in the optical properties, we employed the TB09 meta-GGA\cite{Tran_2009}, which is known to reach almost the GW accuracy\cite{Borlido_2019}.

\begin{figure*}
    \centering
    \includegraphics[width=\textwidth]{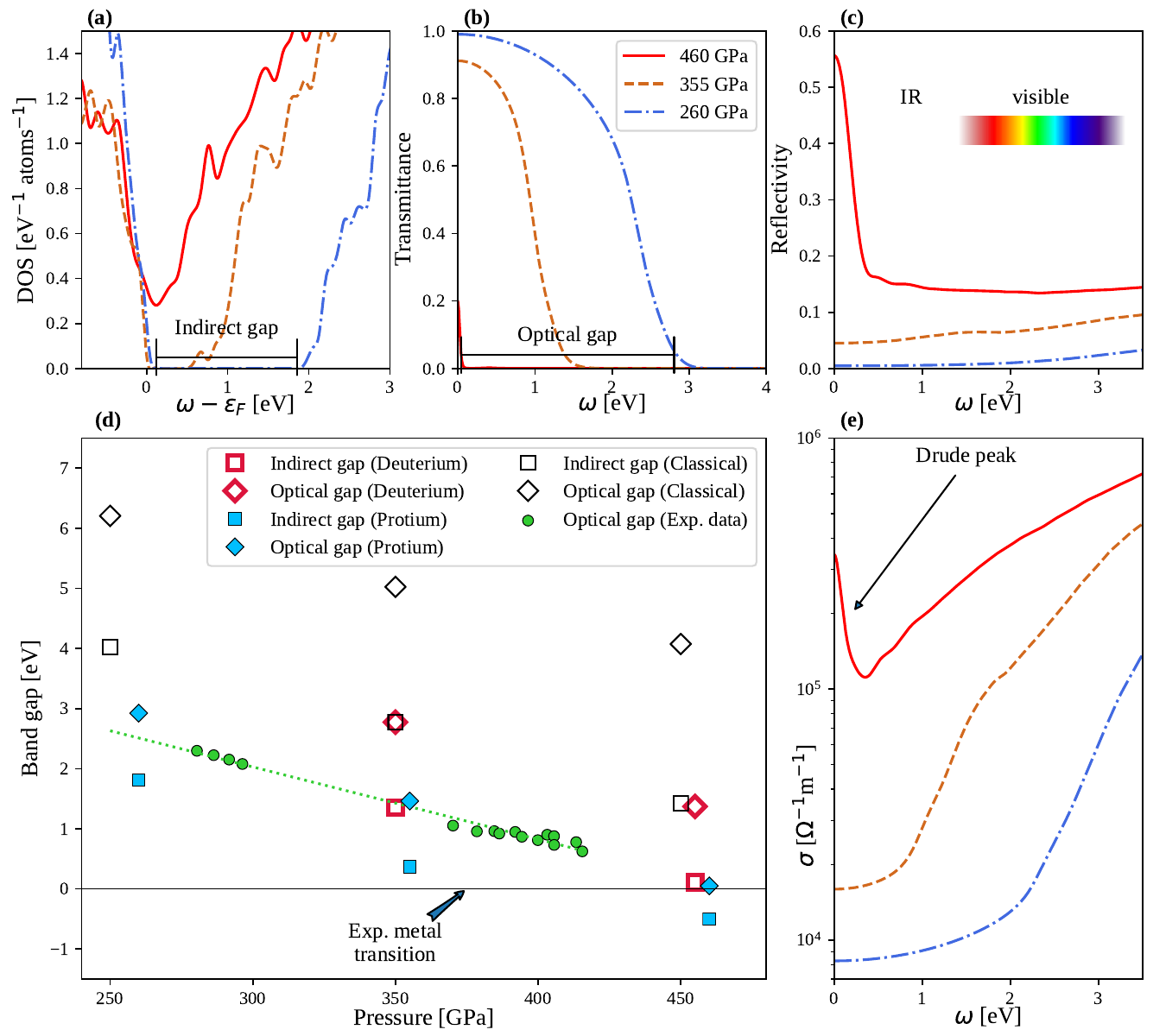}
    \caption{Optical properties of phase III of hydrogen, including quantum nuclear fluctuations, sampled in a supercell with 432 atoms. \textbf{(a)} Electron DOS at three pressures (protium).  The indirect gap is measured as the region where the DOS per atom drops below \SI{0.001}{\per\electronvolt}. \textbf{(b)} Transmittance over a sample of \SI{1.5}{\micro\meter} (protium). The transmittance is used to compute the direct (optical) gap as shown by the bars for the case of \SI{260}{\giga\pascal}. \textbf{(c)} Reflectivity of hydrogen at different pressures (protium). Even if the phase is a metal above \SI{375}{\giga\pascal}, the reflectivity in the IR remains very low (under 20\% at \SI{460}{\giga\pascal}) in the whole visible range (marked with a rainbow). So we predict the phase to be transparent at \SI{250}{\giga\pascal}, black at \SI{355}{\giga\pascal} and remaining black even at higher pressures, where also the optical gap is closed. \textbf{(d)} The indirect and direct gap for protium, deuterium and the classical nuclei (infinite mass), compared with the experimental results\cite{Loubeyre2019observation}. The metallic transition due to indirect gap closure occurs experimentally (protium) at \SI{360}{\giga\pascal} at \SI{200}{\kelvin}\cite{Eremets2019_hydro}. We refer to ``classical'' the gaps computed in the static equilibrium configuration, while for protium and deuterium (hydrogen-1 and 2) the optical properties computed in a snapshot sampled from the quantum nuclei wave-function (see Methods). \textbf{(e)} Optical conductivity (protium). For clarity, in panels (\textbf{b,c,e}) we report the data considering only the electronic contribution to the susceptibility (see Methods). }
    \label{fig:bands}
\end{figure*}

The closure of the gap in the density of states (DOS) coincides with the onset of conductivity. Here this occurs at \SI{380}{\giga\pascal} (see \figurename~\ref{fig:bands}, panels \textbf{a,d}) at \SI{0}{\kelvin}, in good agreement with the conductivity measurements in Ref.\cite{Eremets2019_hydro} (\SI{360}{\giga\pascal} at \SI{200}{\kelvin}). However, after the fundamental gap closes, the Drude peak is small (due to the small value of the DOS), as we show in panel \textbf{(e)}. Thus, the sample remains transparent within an IR window that extends up to the direct gap energy associated with interband optical transitions (panel \textbf{b}). The low DOS around the Fermi level of phase III (panel \textbf{a} of \figurename~\ref{fig:bands}) indicates that this phase will not be a high-temperature superconductor.
The closure of the predicted optical gap matches nicely with the measurements of Ref.\cite{Loubeyre2019observation} (panel \textbf{d}).
Finally, even above the metallic transition and the optical gap closure, reflectivity in the optical range is very flat and small because of interband transitions (panel \textbf{c}).
This gives a dark appearance to the sample, as observed in experiments. 

From the remarkable agreement with experiments, our calculations strongly support the C2/c-24 structure as the phase III also from optical properties, at odds with previous results\cite{Azadi_2017}.
Interestingly, we find quantum fluctuations to have a stronger impact on the optical gap, downshifted by more than \SI{3}{\electronvolt}, while the fundamental indirect gap is downshifted by about \SI{2}{\electronvolt}.
This is probably because indirect gap transitions may contribute to optical transmittance in presence of a strong electron-phonon coupling: an optical photon can excite an electron-hole pair at different reciprocal lattice vectors if also a phonon is emitted in order to keep momentum conservation\cite{Zacharias_2015}. 

In \figurename~\ref{fig:bands} (\textbf{d}) we predict the metallization of deuterium. The indirect gap increases by \SI{1}{\electronvolt} and the optical by about \SI{1.2}{\electronvolt} with respect to protium at \SI{0}{\kelvin}. However, since deuterium has lower phonon energy modes, its band structure is expected to be more affected by temperature than protium. This isotope effect on the electronic properties is the largest we are aware of in a solid crystal.
If no phase-transitions occurs, deuterium should become a black metal at \SI{450}{\giga\pascal}, \SI{70}{\giga\pascal} higher than protium.

Our calculations show that there is no contradiction between the existence of a conducting metallic state, an IR transparent window, and a black appearance of the sample.
As our results predict a low reflectivity at \SI{460}{\giga\pascal}, we rule out the hypothesis of C2/c-24 becoming shiny at very high pressure, and the claimed metallization observed at \SI{490}{\giga\pascal} must be related to a first-order phase-transition\cite{Dias2017}, probably to an atomic metallic phase\cite{Borinaga_2018}.

\vspace*{5mm}
\noindent
In the process of submitting the current paper, a related work appeared on arXiv [arXiv:1911.06135] supporting some of our conclusions on the gap closure in phase III.\\

\noindent
{\sf{\textbf{\large{Acknowledgements}}}}\\
\added[]{
L.M. acknowledges the CINECA award under the ISCRA initiative, for the availability of high performance computing resources and support. This work was performed using HPC resources from Idris and The Grand Challenge Jean Zay. 
I.E. has received funding from the European Research Council (ERC) under the European Unions Horizon 2020 research and innovation program (grant agreement No. 802533).
}\\

\noindent
{\sf{\textbf{\large{Author contributions}}}}\\
The project was conceived by all authors. 
L.M. made the analytical and numerical calculations and prepared
the figures with inputs from all the authors.
All authors contributed to the redaction of the manuscript. \\

\noindent
{\sf{\textbf{\large{Competing interests}}}}\\
The authors declare no competing interests.\\

\noindent
{\sf{\textbf{\large{Additional information}}}}\\
\textbf{Correspondence and request of materials}  should be addressed to F.M.\\

\noindent
    {\sf{\textbf{\large{Methods}}}}\\

To include quantum fluctuations and anharmonicity at a non-perturbative level, we use the stochastic self-consistent harmonic approximation (SSCHA)\cite{Errea2014,Bianco2017,Monacelli2018A}.
The SSCHA is a variational method: it looks for the density matrix that minimizes the Gibbs free energy:
\beq
G(P, T) = \min_{\tilde\rho}\Tr\left[\tilde\rho H + k_bT \tilde\rho \ln \tilde\rho + P \Omega(\tilde \rho) \right],
\label{eq:g:func}
\eeq
where $\Omega(\rho)$ is the crystal volume, $H$ is the ionic Hamiltonian in the Born-Oppenheimer (BO) approximation,
\beq
H = T + V,
\eeq
with $T$ the ionic kinetic energy and $V$ the BO energy landscape. 
The trial density matrix $\tilde\rho$ is chosen among all possible Gaussians.
In this way, the only variational parameters are the average centroid positions and the fluctuations around the average.

The BO energy landscape $V$ is calculated within density functional theory (DFT) using the generalized gradient approximation (GGA) BLYP\cite{BLYP}, as implemented in Quantum ESPRESSO\cite{Giannozzi2009,Giannozzi2017}. 
The static and dynamical SSCHA calculations were performed in a $2\times 2\times 1$ supercell of the primitive unit cell, which consists of 96 atoms. Each supercell configuration was computed with a uniform grid of $4\times 4 \times 4$ for the Brillouin zone integrals, with a wave-function cutoff of 60 Ry (240 Ry for the electronic density). The convergence of results was checked with a $6\times 6\times 6$ grid and a wave-function cutoff of 80 Ry (320 Ry for the density). We generated a norm-conserving pseudo-potential with no pseudized electrons using the settings from the Pseudo Dojo\cite{van_Setten_2018} library and the ONCVPSP software\cite{Hamann_2013}.
In all calculations, Marzari-Vanderbilt smearing of 0.03 Ry was used to account for the enhanced gap closure in the 96 atom cell when the phonon-distorted configuration is considered (see SI).


The indirect band gap is computed from the electron DOS, as shown in \figurename~\ref{fig:bands} (panel \textbf{a}). The DOS is obtained simulating a configuration with 432 atoms ($3\times 3\times 2$ supercell) with ions randomly distributed according to the SSCHA nuclear wave-function. The electronic states are obtained through a DFT calculation with TB09 functional as implemented in Quantum ESPRESSO\cite{Germaneau_2013}, with a Brillouin zone sampling of $8\times 8 \times 6$. Interestingly, we find TB09 to reproduce the fundamental band gap for the ideal crystal (open squares in \figurename~\ref{fig:bands}, panel \textbf{d}) in very good agreement with $G_0W_0$ calculations of Ref.\cite{McMinis_2015} and Quantum Monte Carlo of Ref.\cite{Azadi2017shissor}, with a computational cost reduced of many orders of magnitude.
The direct gap is computed from the simulated transmittance over a sample of $\SI{1.5}{\micro\meter}$, as illustrated in \figurename~\ref{fig:bands} (panel \textbf{b}).
The dielectric tensor and the optical conductivity, 
\begin{equation}
    \bm{\varepsilon}^{(tot)}(\omega) = 1 + 4\pi\bm{\chi}^{(tot)}(\omega)
    \label{eq:dielectric}
\end{equation}
and
\begin{equation}
    \bm{\sigma}^{(tot)}(z) = -i z \bm{\chi}^{(tot)}(z),
\end{equation}
are computed from the total susceptibility. This is divided into an electronic and a ionic contribution:
\begin{equation}
    \bm{\chi}^{(tot)} = \bm{\chi} + \bm{\chi}^{(ion)}.
\end{equation}
All the quantities with $^{(tot)}$ are computed accounting for both the ionic and electronic contribution, otherwise only the electronic contribution is included.
The electronic susceptibility is computed within the independent particle approximation framework on the same phonon-displaced configuration as the DOS. We checked that the results did not change if the operation is repeated in 5 different configurations.
\begin{align}
\chi_{\alpha\beta}(z) =- \frac{4e^2}{m^2N_k\Omega}& \sum_{kmn}^{n\ge m}
\frac{f(\xi_{km}) - f(\xi_{km})}{\xi_{km} - \xi_{kn}} \nonumber \\ &\frac{\braket{km| p_\alpha| kn}\braket{kn| p_\beta|km}}{(\varepsilon_{km} - \varepsilon_{kn})^2 - z^2},
\label{eq:chi:good}
\end{align}
\begin{equation}
    \xi_{km} = \varepsilon_{km} - \varepsilon_f\qquad z = \omega + i \eta,
\end{equation}
where $N_k$ is the number of $k$ points used in the sum, $\Omega$ the simulation cell volume, $f$ the Fermi occupation function, $\varepsilon_{kn}$ and $\varepsilon_f$, respectively, the energy of the state $\ket{kn}$ and the Fermi energy, $\eta$ the smearing, and $\braket{km|p_\alpha|kn}$ is the momentum matrix element along the Cartesian direction $\alpha$ for the optical transition between the $m$ and $n$ states at the $k$ point in the reciprocal space. \eqname~\eqref{eq:chi:good} includes both interband and intraband terms, enabling to correctly simulate the disordered phonon displaced configurations in the supercell even when interband and intraband transitions are not well defined. We implemented \eqname~\eqref{eq:chi:good} into the epsilon.x code of Quantum Espresso.
We used a smearing $\eta$ of $\SI{0.1}{\electronvolt}$ (this is the reason for the non zero conductivity of the insulating phase of \figurename~\ref{fig:bands}). To account for all possible orientations of the crystal we take
\begin{equation}
    \epsilon(\omega) = \frac 1 3\sum_{\alpha= x,y,z} \epsilon_{\alpha\alpha}(\omega).
\end{equation}
The refractive index of the material is
\begin{equation}
    n(\omega) = \sqrt{\varepsilon(\omega)}.\label{eq:n}
\end{equation}
The reflectivity of \figurename~\ref{fig:bands} (panel \textbf{c}) is computed as
\begin{equation}
    R = \left| \frac{n(\omega) - n_d}{n(\omega)  + n_d}\right|^2,
\end{equation}
where $n_d$ is the diamond anvil cell refractive index, that is assumed constant in the simulated interval to 2.33.
The transmittance across a sample of thickness $d=\SI{1.5}{\micro\meter}$ is simulated as
\begin{equation}
    T = (1 - R)^2\exp\left(- 2 \omega \Im n \frac{d}{c}\right),
\end{equation}
(we neglected multiple reflections inside the sample).
The optical band gap is considered as the first value for which the transmittance drops below 2\% (excluding the absorption due to the vibrational modes).
In \figurename~\ref{fig:bands}, the ionic contribution to conductivity, transmittance, and reflectivity was neglected. However, it is important to simulate the vibrational spectrum. The ionic susceptibility is the dipole-dipole correlation function:

\begin{equation}
    \chi_{\alpha\beta}^{(ion)}(\omega) = \int_{-\infty}^{\infty}e^{-i\omega t}\braket{M_\alpha(t)M_\beta(0)} \, dt .
    \label{eq:chiion}
\end{equation}
The average $\braket{\cdot}$ must be performed on the quantum nuclear ground state.
The dipole induced by a ionic displacement is approximated as linear:
\beq 
M_\alpha(t) = |e|\sum_{b = 1}^{3N}Z_{\alpha b}[R_b(t) - R^{(0)}_b],
\label{eq:dipole}
\eeq 
where $Z_{\alpha b}$ is the Born effective charge, $R_b(t)$ is the position operator of the $b$ atom and $R^{(0)}_b$ is the average position of the $b$ atom ($b$ runs over both atomic and Cartesian indices).
In \figurename~\ref{fig:irramspec} (panels \textbf{a,b}), we report the difference of the transmittance of the sample with and without the ionic contribution. This quantity correctly accounts for electrostatic effects neglected by just considering the imaginary part of the susceptibility. In particular, the sample absorbs where the refractive index (\eqname~\ref{eq:n}) has an imaginary part. When the ionic contribution is negligible with respect to the electronic one, the complex refractive index may be approximated as:
\begin{equation}
    n^{(tot)}(\omega) \approx n + \frac{2\pi}{n}\chi^{(ion)} \qquad 4\pi\chi^{(ion)} \ll 1 + 4\pi \chi
\end{equation}
For a transparent material (i.e. the imaginary part of $n$ is zero: $\Im n = 0$), the absorption is directly related to the imaginary part of the ionic susceptibility:
\begin{equation}
    \Im n^{(tot)}(\omega) \approx \frac{2\pi}{n} \Im \chi^{(ion)}(\omega)
\end{equation}
However, in the presence of strong effective charges, the ionic susceptibility can be greater than the electronic one. In this case also the real part of $\chi^{(ion)}$ gives a contribution to the absorption. In particular, $\varepsilon$ becomes negative between the LO and the TO frequencies. This induces absorption (see \eqname~\eqref{eq:n}) in a finite region even in the presence of a phonon with an infinite lifetime. This is a dominant contribution to the IR vibron at \SI{355}{\giga\pascal} (see extended data).

The Raman spectrum is assumed proportional to the polarizability correlation function:
\beq
I_{Raman}(\omega)\propto \Im \int_{-\infty}^\infty dt e^{-i\omega t}\braket{\bm{\alpha}(t)\bm{\alpha}(0)} \label{eq:Raman},
\eeq 
\beq 
\alpha_{ab}(t) =\sum_{c = 1}^{3N} A_{\alpha \beta c}[R_c(t) - R^{(0)}_c],
\label{eq:polar}
\eeq 
here, $A$ is the Raman tensor. The position operator in both \eqname~\eqref{eq:dipole} and \eqref{eq:polar} can be written using phonon creation-annihilation operators. The resulting phonon correlation functions of \eqname~\eqref{eq:chiion} and \eqname~\eqref{eq:Raman} are computed using the SSCHA dynamical Green function\cite{Bianco2017}, accounting for phonon-phonon interactions non perturbatively. For the IR we get:
\begin{equation}
    \chi_{\alpha\beta}^{(ion)}(\omega) = e^2 \sum_{ab}\frac{Z_{\alpha a} Z_{\beta b}}{\sqrt {m_am_b}} G_{ab}(\omega),
    \label{eq:chi:ion:good}
\end{equation}
\begin{equation}
    \frac{G_{ab}(t)}{\sqrt{m_am_b}} = \braket{[R_a(t) - R_a^{(0)}] [R_b(0) - R_b^{(0)}]} ,
\end{equation}
\begin{equation}
    G_{ab}^{-1}(\omega) = \omega^2 - D_{ab} - \Pi_{ab}(\omega) ,
    \label{eq:green}
\end{equation}
$\bm D$ is the SCHA dynamical matrix, $\bm\Pi$ is the phonon self-energy (Ref. \cite{Bianco2018} equation A12), $m_a$ is the atomic mass of the $a$-th atom.
\eqname~\eqref{eq:green} has been computed using a Lanczos algorithm that will be discussed elsewhere in detail. Noten \eqname~\eqref{eq:chi:ion:good} we trace the Green function $G_{ab}(\omega)$ contracted with the effective charges and we keep both the real and imaginary part to study the absorption, after summing the electronic contribution to it (\eqname~\ref{eq:chi:good}).
For the Raman calculation, we instead take the imaginary part of the response function:
\begin{equation}
    I_{Raman}(\omega) \propto -\sum_{\alpha=1}^3\sum_{ab} \frac{A_{\alpha \alpha a}A_{\alpha\alpha b}}{\sqrt{m_a m_b}} \Im G_{ab}(\omega).
\end{equation}
Here, we consider only the Raman contribution arising by incoming and outcoming radiation with the same polarization vector.
Also in this case, the Green function is contracted with the Raman tensor.

In the self-energy expression $\bm\Pi(\omega)$, we compared the results with and without the 4 phonon scattering vertex (the $\stackrel{(4)}{\bm\Phi}$ of Ref. \cite{Bianco2018} equation A12), obtaining no significative difference. Thus, all the simulations have been performed setting $\stackrel{(4)}{\bm\Phi} = 0$ ($\bm\Pi(\omega)$ reduces to equation A14 of Ref.\cite{Bianco2018}).
The effective charges and the Raman tensor of \eqname~\eqref{eq:dipole} and \eqref{eq:polar} are computed with Quantum ESPRESSO phonon packages in the SSCHA average centroid position.
The Raman tensor was computed within the LDA approximation\cite{Lazzeri_2003}.
For the \SI{460}{\giga\pascal} calculation, we computed the Raman tensor and effective charges forcing the system to be an insulator, thus our results are indicative. A more sophisticated time-dependent approach should be used for metals to correctly describe them\cite{Bistoni2019}.

\bibliographystyle{naturemag}
\bibliography{biblio}

\end{document}